\documentclass[fleqn,usenatbib]{mnras}

\usepackage{amsfonts}

\usepackage[T1]{fontenc}
\usepackage{ae,aecompl}
\usepackage[position=top]{subfig}
\usepackage{float}
\usepackage{comment}

\usepackage{grffile}
\usepackage{etoolbox}
\makeatletter 
  \patchcmd{\NAT@citex}
    {\@citea\NAT@hyper@{%
      \NAT@nmfmt{\NAT@nm}%
      \hyper@natlinkbreak{\NAT@aysep\NAT@spacechar}{\@citeb\@extra@b@citeb}%
      \NAT@date}}
    {\@citea\NAT@nmfmt{\NAT@nm}%
    \NAT@aysep\NAT@spacechar\NAT@hyper@{\NAT@date}}{}{}

  \patchcmd{\NAT@citex}
    {\@citea\NAT@hyper@{%
      \NAT@nmfmt{\NAT@nm}%
      \hyper@natlinkbreak{\NAT@spacechar\NAT@@open\if*#1*\else#1\NAT@spacechar\fi}%
        {\@citeb\@extra@b@citeb}%
      \NAT@date}}
    {\@citea\NAT@nmfmt{\NAT@nm}%
    \NAT@spacechar\NAT@@open\if*#1*\else#1\NAT@spacechar\fi\NAT@hyper@{\NAT@date}}
    {}{}
\makeatother

\usepackage{etoolbox}
 \makeatother
\usepackage{graphicx}	
\usepackage{amsmath}	
\usepackage{amssymb}	
\usepackage{pifont}
\newcommand{\uH}{\mathrm{H}}

\title[Lessons on early star formation from XLSSC~122]{Lessons on early structure formation from a mature galaxy cluster observed at cosmic noon}

\author[B. Liu, A. Schauer, V. Bromm]{Boyuan Liu\textsuperscript{\href{https://orcid.org/0000-0002-4966-7450}{\includegraphics[width=2.5mm]{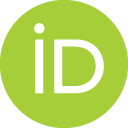}}\,}\thanks{E-mail: boyuan@utexas.edu}$^{1}$, Anna
T. P. Schauer\thanks{Hubble Fellow}$^{1}$, and Volker Bromm$^{1}$
\\
$^{1}$Department of Astronomy, University of Texas, Austin, TX 78712, USA\\
}

\date{Accepted XXX. Received YYY; in original form ZZZ}

\pubyear{2020}

\begin{document}
\label{firstpage}
\pagerange{\pageref{firstpage}--\pageref{lastpage}}
\maketitle

\begin{abstract}
We demonstrate a new approach of indirectly constraining both early star and structure formation via mature galaxy clusters at cosmic noon ($z\sim2$), using the cluster XLSSC~122 as an example. With the standard Press Schechter formalism, we infer a rapid evolution of the 
star formation efficiency (the ratio of stellar to halo mass) from $10^{-4}$ to $0.01$ during $z\sim 20-13$, based on the age distribution of stars in post-starburst galaxies of XLSSC~122, measured by HST photometry assuming no dust extinction. Here, we consider all low-mass haloes, including minihaloes, that host the first stars and galaxies ($5\times 10^5\ \rm M_{\odot}\lesssim M_{\rm halo}\lesssim 10^{10}\ \rm M_{\odot}$). We also place new constraints on fuzzy dark matter models of $m_{\rm a}\lesssim 5\times 10^{-21}\ \mathrm{eV}/c^{2}$ for the ultra-light boson mass, from the abundance of galaxies with star formation at $z\gtrsim 13$ in XLSSC~122. Our exploratory results are consistent with existing constraints. More comprehensive results will be obtained if our approach is extended to a large sample of clusters or field post-starburst galaxies at cosmic noon, with improved modelling of halo and stellar populations.
\end{abstract}



\section{Introduction}
\label{s1}
A grand challenge for modern astrophysics is to constrain star and galaxy formation in the first billion years of cosmic history \citep{loeb2013}. 
This is in particular the case for the first, so-called Population~III (Pop~III), stars and galaxies formed at $z\gtrsim 10$, which are believed to have distinct features compared with their present-day counterparts (reviewed by e.g. \citealt{bromm2011first,bromm2013}). Future facilities, such as the {\it James Webb Space Telescope (JWST)} and the Einstein Telescope (ET), are expected to directly probe this formative epoch (e.g. \citealt{appleton2009,pawlik2011first,zackrisson2016spectral,jeon2019signature,basu2019cooking,maggiore2020science}). However, before they come into operation, we have to rely on indirect observations and empirical constraints. 

A traditional indirect approach is `stellar archaeology', where clues to the earliest star-forming environments are derived from local ($z\sim 0$) observations of extremely metal poor (EMP) stars ($\rm [Fe/H]<-3$; e.g. \citealt{frebel2015near,ji2015preserving}). For instance, the non-detection of metal-free stars in the Milky Way place constraints on the low-mass end ($\lesssim 0.8\ \rm M_{\odot}$) of the Pop~III initial mass function (IMF) \citep{hartwig2015, magg2019observational}, while the abundance patterns of observed EMP stars constrain the higher-mass ($\gtrsim 10\ \rm M_{\odot}$) regime of the Pop~III IMF, as well as the properties of the first supernovae \citep{ishigaki2018initial}. Another probe that has become promising recently is the 21-cm signal from high-$z$ neutral hydrogen, encoding the Lyman-$\alpha$ and X-ray fields powered by the first stars and galaxies (e.g. \citealt{Mirocha2019,fialkov2019,21cm2020}). As a specific example,  \citet{schauer2019constraining} found that Pop~III star formation in minihaloes ($M_{\rm halo}\sim 10^{6}\ \rm M_{\odot}$) is required to explain the timing ($z\sim 17$) of the 21-cm absorption signal potentially detected by the Experiment to Detect the Global Epoch of Reionization Signature (EDGES; \citealt{nature21cm}). Each approach is subject to uncertainties, such as turbulent metal mixing in the formation pathways of EMP stars, or the escape fraction of Lyman-$\alpha$ photons. 
It is therefore important to harness new probes to complement the existing ones. 

In this work, we consider mature galaxy clusters at cosmic noon ($z\sim 2$) to constrain early star and structure formation, e.g. IDCS~J1426.5+3508 ($z=1.75$; \citealt{stanford2012idcs}), JKCS~041 ($z=1.8$; \citealt{andreon2014jkcs,newman2014spectroscopic}), Cl~J1449+0856 ($z=2$; \citealt{gobat2013wfc3,strazzullo2016red}), and XLSSC~122 ($z=1.98$; \citealt{mantz2018xxl,willis2020spectroscopic}). These clusters are the most massive virialized structures in the Universe $\sim 3-4\ \mathrm{Gyr}$ after the Big Bang, such that their oldest constituent post-starburst galaxies (i.e. red-sequence cluster members) probe star formation histories at much earlier times ($z\gtrsim 10$). Here, we consider XLSSC~122 to demonstrate our basic approach, given the cluster's clear post-starburst red-sequence members with ages of $t_{\rm w}\sim 2.4-3.1\ \rm Gyr$. We construct an idealized stellar population model, based on the standard extended Press-Schechter (EPS) formalism (Sec.~\ref{s2}), to derive new constraints on early star formation parameters and dark matter (DM) physics from the age distribution of stars within the XLSSC~122 red sequence (Sec.~\ref{s3}), as measured by the Hubble Space Telescope (HST; \citealt{willis2020spectroscopic}). We summarize our findings and discuss future directions in Section~\ref{s4}.

\section{Stellar Population Model}
\label{s2}
The cluster XLSSC~122 is observed at $z_{\rm obs}=1.98$ \citep{willis2020spectroscopic}. X-ray observations show that it has a virial mass of $M_{2}\sim 10^{14}\ \mathrm{M_{\odot}}$, a virial (physical) radius of $r_{200}\approx 1.5 r_{500}\simeq 440\ \mathrm{kpc}$, and a sound-crossing time of $t_{\mathrm{cr}}\simeq 3.3\times 10^{8}\ \mathrm{yr}$ \citep{mantz2018xxl}. To connect the observed luminosity-weighted posterior age distribution of (red-sequence) galaxies in XLSSC~122 to star and structure formation at higher redshifts ($z\gtrsim 13$) with a simple but flexible model, we employ the standard EPS formalism \citep{mo2010galaxy} to estimate halo abundances. We further make the following assumptions and approximations:

\begin{itemize}
\item XLSSC was formed at $z_{2}\simeq 2.8$, i.e. 3 sound-crossing timescales before the observed epoch ($t_{\mathrm{cr}}\simeq 3.3\times 10^{8}\ \mathrm{yr}$)\footnote{Hydrodynamical simulations of gas in a forming cluster indicate that virial equilibrium is achieved within a minimum of 2 to
3 sound-crossing timescales \citep{roettiger1998anatomy}. We adopt $\Delta t_{\rm obs}=3t_{\rm cr}$ as a conservative estimate of the delay time between cluster formation and observation. In general, lower $\Delta t_{\rm obs}$ leads to higher HSFE, but the variation is minor (within 30\%) for $\Delta t_{\rm obs}\lesssim 3t_{\rm cr}$. }. 
\item Star formation in progenitor haloes older than $t_{\rm w,peak}\simeq 2.98$~Gyr (corresponding to formation at $z_{1}\gtrsim 13$) is unaffected by environmental effects/cosmic variance, and reflects the average star formation efficiency in the early Universe.
\item In such high-$z$ haloes, star formation is episodic on timescales smaller than the age distribution bin size of $\sim 50$~Myr, such that the luminosity-weighted posterior age distribution of galaxies in XLSSC~122 is a good approximation to the underlying stellar age distribution, assuming a universal mass-to-light ratio.
\end{itemize}

The observational input comes from the last four bins in the stellar age distribution of XLSSC~122 assuming no dust absorption ($A_{V}=0$, see fig.~4 of \citealt{willis2020spectroscopic}), corresponding to the four redshift bins: $z_{1}\sim 12.6-13.6$, $13.6-14.8$, $14.8-16.3$ and $16.3-18.4$ (with \textit{Planck} cosmological parameters, see Sec.~\ref{s3.1}), contributing $\simeq 0.275$, 0.124, 0.025 and 0.004 of the total stellar mass/luminosity, respectively. Here we focus on the $A_{V}=0$ model, as it predicts the oldest stellar ages (up to $z\sim 18$), most relevant for the first galaxies and stars. We build a simple stellar population model for XLSSC~122 with minimum parameters, as described below. With this model, information on high-$z$ star formation and DM physics can be extracted by matching the observed stellar mass (in galaxies) within a given age range to the model predictions.

\begin{figure}
    \centering
    \includegraphics[width=\columnwidth]{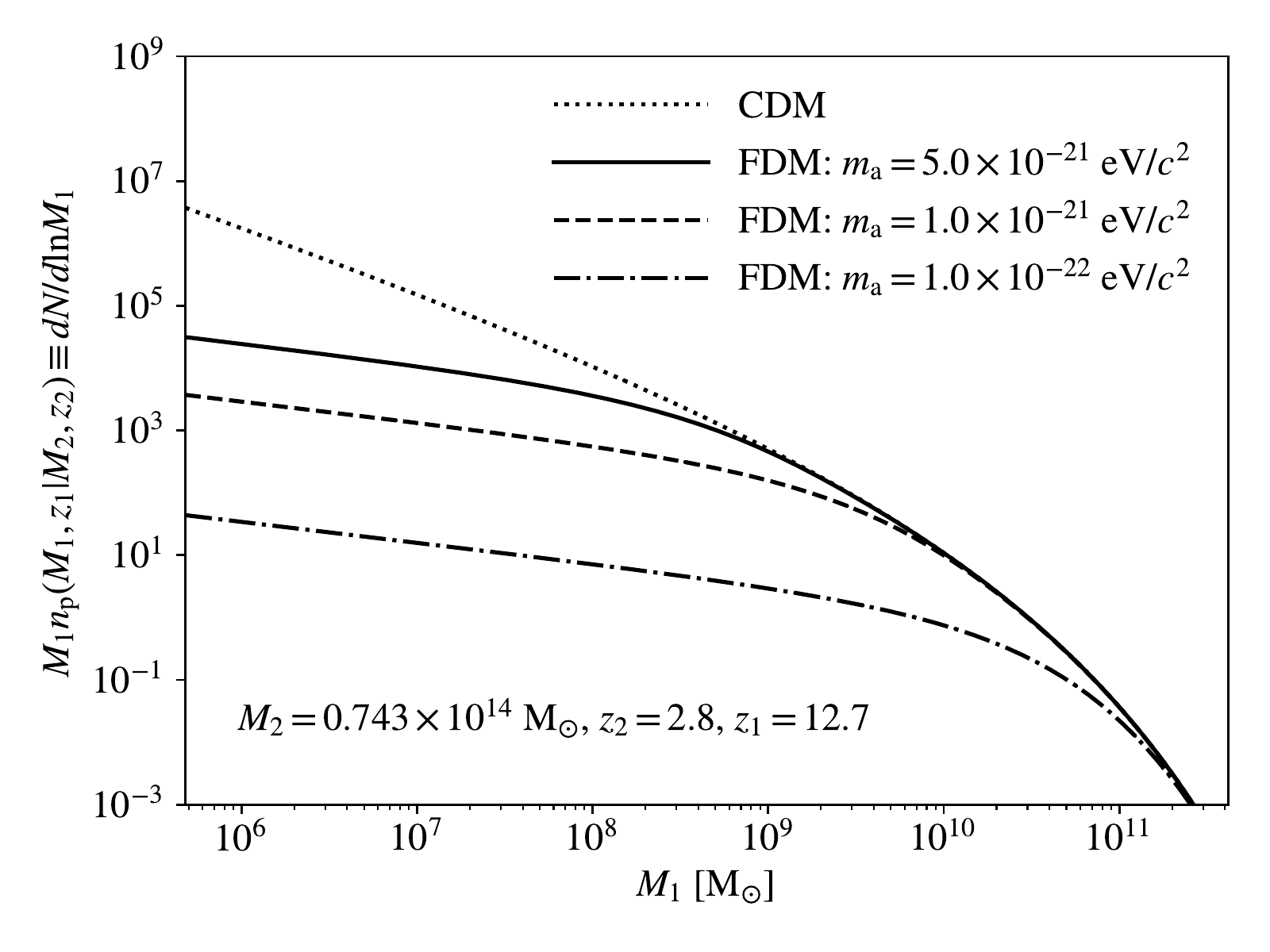}
    \caption{Progenitor mass functions of XLSSC~122 at $z_{1}=12.7$, for the standard $\Lambda$CDM (dotted), and FDM models (see Sec.~\ref{s3.3}) with $m_{\mathrm{a}}c^{2}=5\times 10^{-21}\ \mathrm{eV}$ (solid), $10^{-21}\ \mathrm{eV}$ (dashed) and $10^{-22}\ \mathrm{eV}$ (dash-dotted). Here we assume that XLSSC~122 (observed at $z_{\mathrm{obs}}\simeq1.98$) has a virial mass of $M_{2}\sim 10^{14}\ \mathrm{M_{\odot}}$, and formed 3 sound crossing timescales ago ($t_{\mathrm{cr}}\simeq 3.3\times 10^{8}\ \mathrm{yr}$) at $z_{2}\simeq 2.8$, based on X-ray observations \citep{mantz2018xxl}. }
    \label{f4}
\end{figure}

Under the episodic star formation assumption, the mass of stars formed in haloes within the mass range $M_{1}\sim M_{1}+\delta M_{1}$ and redshift bin $i$ [$z_{1,i}-\Delta z_{1,i}/2; z_{1,i}+\Delta z_{1,i}/2$] is 
\begin{align}
\delta M_{\star,i}=\eta(z_{1,i},M_{1})\Delta n_{\rm p}(z_{1,i},M_{1})M_{1}\delta M_{1}\ ,\label{e1}
\end{align}
where $\eta\equiv\langle \Delta M_{\star}\rangle/\Delta M_{\rm halo}=\eta(z,M_{\rm halo})$ is the \textit{instantaneous} halo star formation efficiency (HSFE, the average mass of newly-formed stars per increase in halo mass), and $\Delta n_{\rm p}(z_{1,i},M_{1})$ is the number of progenitor haloes of XLSSC~122 per unit halo mass in the mass range [$M_{1}; M_{1}+\delta M_{1}$], formed in redshift bin $i$. For simplicity, we estimate $\Delta n_{\rm p}(z_{1,i},M_{1})$ with
\begin{align}
\Delta n_{\rm p}(z_{1,i},M_{1})&=n_{\rm p}(z_{1,i}-\Delta z_{1,i}/2, M_{1}|z_{2},M_{2})\notag\\
&-n_{\rm p}(z_{1,i}+\Delta z_{1,i}/2, M_{1}|z_{2},M_{2})\ ,\label{e2}
\end{align}
where 
$n_{\rm p}(z_{1},M_{1}|z_{2},M_{2})\equiv dN/dM_{1}$ is the cluster progenitor mass function\footnote{$n_{\rm p}(z_{1},M_{1}|z_{2},M_{2})\delta M_{1}\equiv$ number of haloes in the mass range $[M_{1}; M_{1}+\delta M_{1}]$ at $z_{1}$ that end up in a halo of a mass $M_{2}$ at $z_{2}$.} at $z_{1}$, which only depends on cosmology, reflected in the linear power spectrum $P(k)$. Here, we calculate the progenitor mass functions with the standard EPS formalism, without imposing any corrections based on cosmological simulations, which allows us to take into account different cosmologies self-consistently. Fig.~\ref{f4} shows four examples of $n_{\rm p}$ for the standard lambda cold dark matter ($\Lambda$CDM) cosmology and three fuzzy dark matter (FDM) models (see Sec.~\ref{s3.3} for details).

\begin{figure}
    \centering
    \includegraphics[width=\columnwidth]{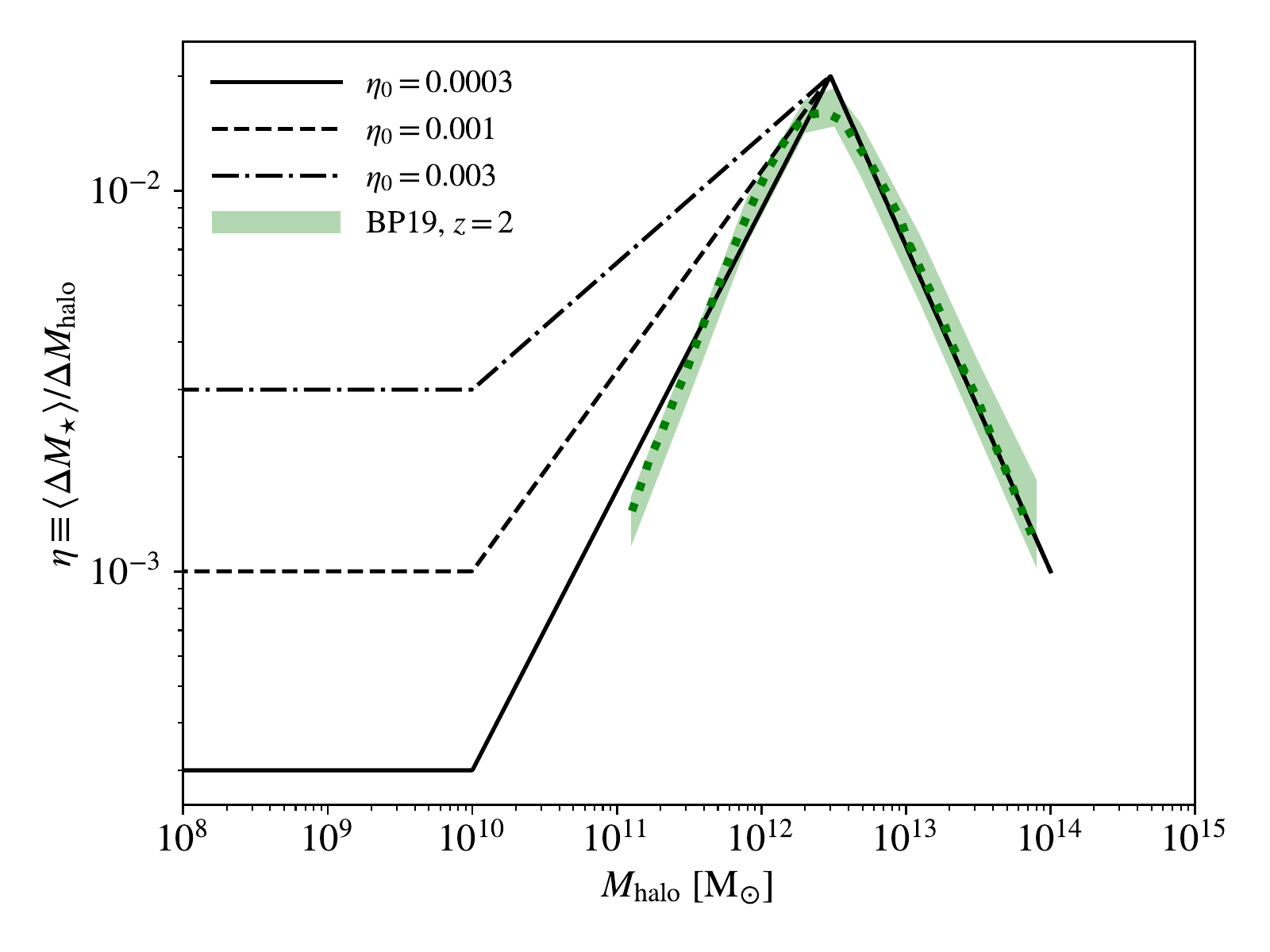}
    \caption{Halo star formation efficiency ($\eta\equiv \langle \Delta M_{\star}\rangle/\Delta M_{\mathrm{halo}}$) models for $\eta_{0}=3\times 10^{-4}$ (solid), 0.001 (dashed) and 0.003 (dash-dotted). Here we assume a constant $\eta=\eta_{0}$ for $M_{\mathrm{halo}}<M_{\mathrm{low}}=10^{10}\ \mathrm{M_{\odot}}$, and power laws between $M_{\mathrm{low}}$, the peak mass $M_{\mathrm{peak}}=3\times 10^{12}\ \mathrm{M_{\odot}}$ and the high-mass reference $M_{\mathrm{high}}=10^{14}\ \mathrm{M_{\odot}}$. For simplicity, we set $\eta(M_{\mathrm{peak}})=0.02$ and $\eta(M_{\mathrm{high}})=0.001$, independent of redshift, based on abundance matching results at $z= 2$ in  \citealt{behroozi2019universemachine} (BP19), shown with the green thick-dotted curve and shaded region for the 68\% confidence interval (see their figs.~9 and 10). }
    \label{f5}
\end{figure}

We parameterize $\eta(M_{1})$ as follows: (i) $\eta=\eta_{0}$ is constant for $M_{1}<M_{\mathrm{low}}=10^{10}\ \mathrm{M_{\odot}}$, and (ii) $\eta$ exhibits (broken) power-law behavior between $M_{\mathrm{low}}$, the peak mass $M_{\mathrm{peak}}=3\times 10^{12}\ \mathrm{M_{\odot}}$, and the high-mass reference point $M_{\mathrm{high}}=10^{14}\ \mathrm{M_{\odot}}$ (see Fig.~\ref{f5} for examples). We set $\eta(M_{\mathrm{peak}})=0.02$ and $\eta(M_{\mathrm{high}})=0.001$, independent of redshift, based on the abundance matching results at $z= 2$ \citep{behroozi2019universemachine}. Here we assume that $\eta$ is constant at the low-mass end for simplicity, in agreement with the semi-analytical analysis of \citet{Mirocha2019} to explain the EDGES 21-cm absorption signal \citep{nature21cm}. Note that at $z_{1}\gtrsim 13$, haloes with $M_{1}\lesssim 10^{10\ (8)}\ \mathrm{M_{\odot}}$ contribute $\gtrsim 99.9\ (90)\%$ of the total mass of progenitor haloes, such that the behavior of $\eta$ at $M_{1}\gtrsim 10^{10}\ \mathrm{M_{\odot}}$ is unimportant for star formation at such high redshifts, i.e. $\eta\approx \eta_{0}$ for $z_{1}\gtrsim 13$. Nevertheless, $\eta$ at the high-mass end is important for determining the total stellar mass $M_{\star,\mathrm{tot}}$ in XLSSC~122 at $z_{\mathrm{obs}}\approx 2$. In our case, $M_{\star,\mathrm{tot}}=M_{2}\eta(z=2,M_{2})\simeq 10^{11}\ \mathrm{M_{\odot}}$, according to \citet{behroozi2019universemachine}. 

Finally, given equations~(\ref{e1}) and (\ref{e2}), observation and theory is bridged with
\begin{align}
f_{i}M_{\star,\mathrm{tot}}=\int_{M_{\mathrm{th}}}^{M_{2}}\Delta n_{\rm p}(z_{1,i},M_{1})M_{1}\eta(z_{1,i},M_{1})dM_{1}\ .\ \label{e3}
\end{align}
On the left-hand side (observation), $f_{i}$ is the fraction of stars formed at redshift bin $i$, while on the right side (theory), there are two unknown/degenerate star formation parameters: $\eta_{0}(z_{1,i})$ and $M_{\mathrm{th}}$, which is the minimum halo mass for star formation. With 
equation~(\ref{e3}), constraints on $M_{\mathrm{th}}$ and $\eta_{0}(z_{1,i})$ can be derived for any given cosmology embodied by the progenitor mass function $n_{\rm p}(z_{1},M_{1}|z_{2},M_{2})$.

We have thus connected the observed age distribution of galaxies/stars to two key theoretical ingredients: cosmology ($n_{\rm p}$) and star formation model ($\eta_{0}$ and $M_{\rm th}$), which are degenerate to some extent. In a full Bayesian model, the joint posterior distributions of model parameters could be derived once the statistical properties of observational data and priors are available. We defer such complex treatment to future work. Instead, in the following section, we explore the bounds on individual parameters separately, by fixing a subset of them to existing constraints.

\section{Constraints on star formation and dark matter physics}
\label{s3}
\subsection{Cold dark matter}
\label{s3.1}

\begin{figure}
    \centering
    \includegraphics[width=\columnwidth]{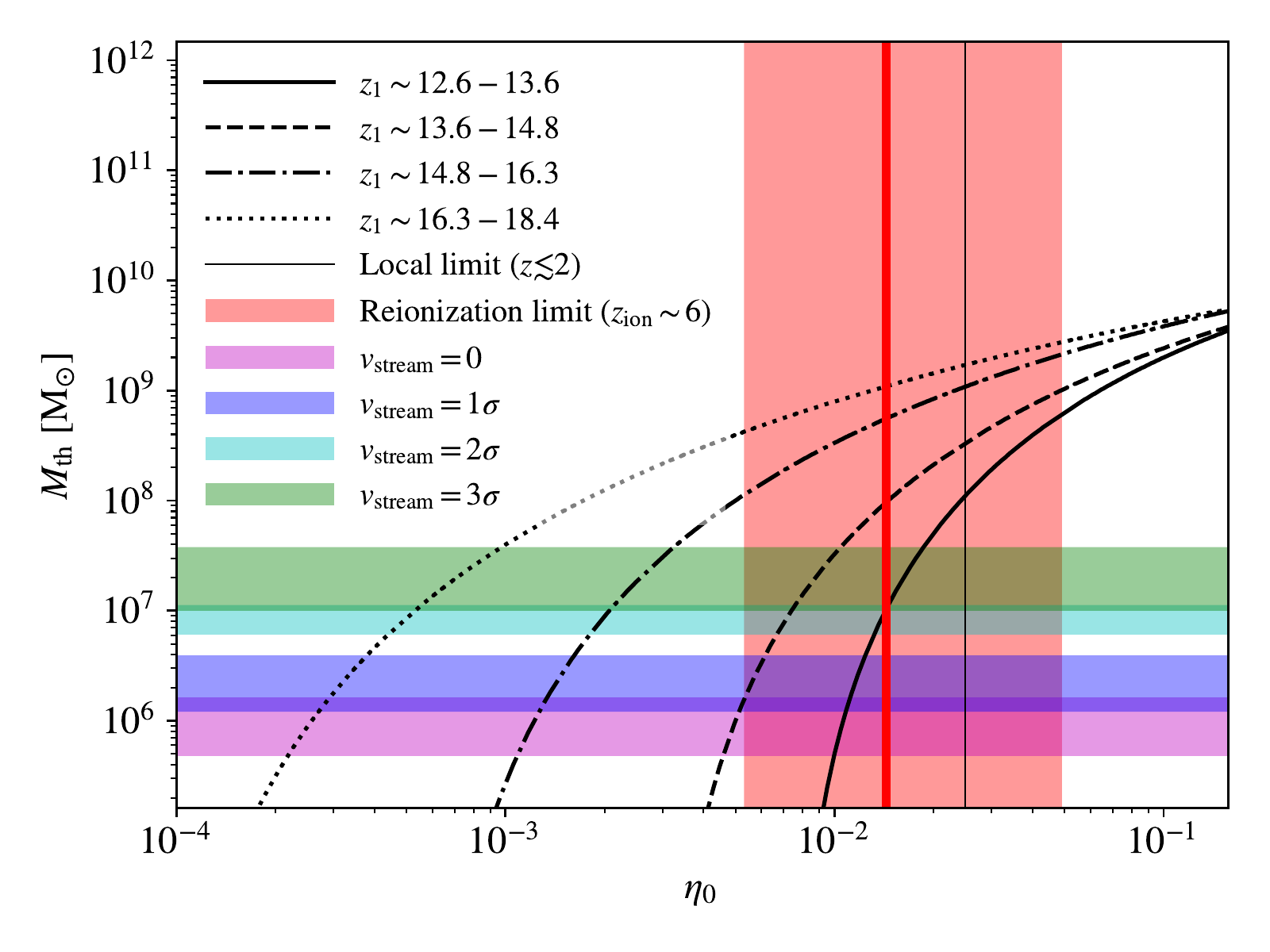}
    \caption{Star formation threshold masses as functions of $\eta_{0}$ for redshift bins $z_{1}\sim 12.6-13.6$ (solid), $1.36-14.8$ (dashed), $14.8-16.3$ (dash-dotted) and $16.3-18.4$ (dotted). Threshold masses for Pop~III star formation from the simulations of \citet{Anna2018} are shown with purple, blue, cyan and green shaded regions for 0, $1\sigma$, $2\sigma$ and $3\sigma$ streaming velocities, respectively. The upper limits on $\eta_{0}$ from reionization (see Sec.~\ref{s3.2}) at $z_{\rm ion}= 6$ with $\hat{f}_{\rm esc}=0.3$ \citep{so2014fully} and abundance matching based on local ($z\lesssim2$) observations \citep{behroozi2019universemachine} are shown with the thick and thin vertical lines. For the former, the range for $\hat{f}_{\rm esc}\sim 0.1-0.7$ is shown with the red shaded region.}
    \label{f1}
\end{figure}

To begin with, we apply equation~(\ref{e3}) to the standard $\Lambda$CDM cosmology with \textit{Planck} parameters: $\Omega_{\rm m}=0.3089$, $\Omega_{\rm b}=0.0486$, $H_{0}=67.74\ \mathrm{km\ s^{-1}\ Mpc^{-1}}$, $\sigma_{8}=0.8159$, $n_{\rm s}=0.9667$, and $N_{\rm eff}=3.046$ \citep{planck}. The corresponding linear power spectrum $P_{\rm CDM}(k)$ is obtained from the \textsc{python} package \textsc{colossus}\footnote{\url{https://bdiemer.bitbucket.io/colossus/}} \citep{colossus}. 
The resulting constraints on $M_{\mathrm{th}}$ and $\eta_{0}$ are shown in Fig.~\ref{f1}, represented by the curves for different redshift bins in $M_{\rm th}-\eta_{0}$ space. Generally, the observed stellar mass of a certain age (corresponding to a given redshift bin) is more sensitive to $\eta_{0}$ than $M_{\rm th}$ (for $v_{\rm stream}\lesssim 3\sigma$), such that 1 dex variation in $M_{\rm th}$ corresponds to less than 0.5 dex variations in $\eta_{0}$, especially for $z_{1}\lesssim 15$.

\begin{figure}
    \centering
    \includegraphics[width=\columnwidth]{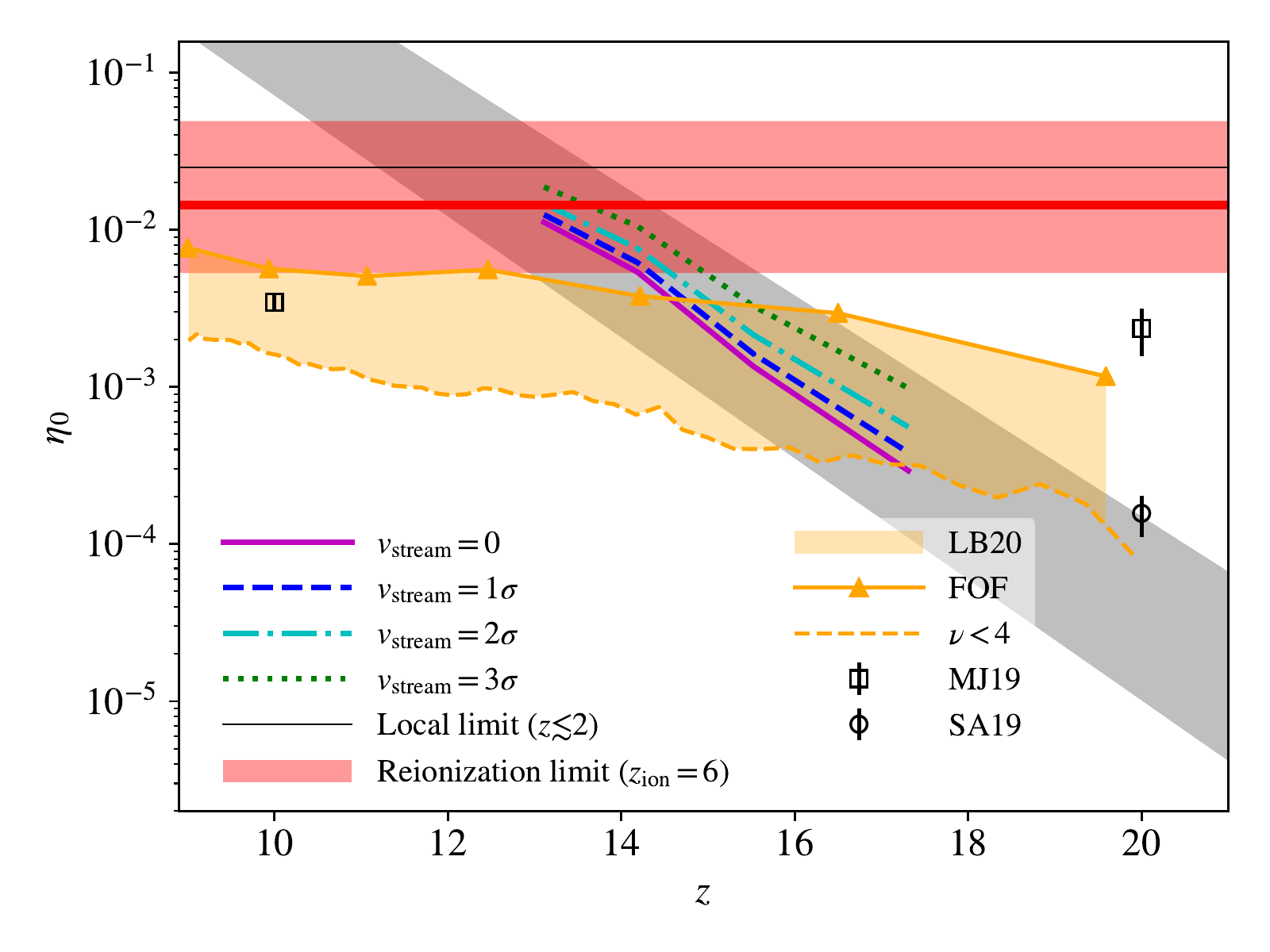}
    \caption{High-$z$ star formation efficiency vs. $z$, inferred from the stellar age distribution of XLSSC~122 under the $A_{V}=0.0$ model \citep{willis2020spectroscopic}, for 0 (solid), $1\sigma$ (dashed), $2\sigma$ (dash-dotted) and $3\sigma$ (dotted) streaming velocities. The $1\sigma$ results are fitted to $\eta_{0}\simeq 2.7\times 10^{-3-0.37(z-15)}$, shown with the gray shaded region (with standard uncertainties). Upper limits on $\eta_{0}$ from reionization (see Sec.~\ref{s3.2}) at $z_{\rm ion}= 6$ with $\hat{f}_{\rm esc}=0.3$ \citep{so2014fully} and abundance matching based on local ($z\lesssim2$) observations \citep{behroozi2019universemachine} are shown with the thick and thin horizontal lines. For the former, the range for $\hat{f}_{\rm esc}\sim 0.1-0.7$ is shown with the red shaded region. For comparison, we show the estimated $\eta_{0}$ from the EDGES 21-cm absorption signal \citep{nature21cm} in the semi-analytical models of \citealt{schauer2019constraining} (SA19; empty circle) and \citealt{Mirocha2019} (MJ19; empty squares), with and without special consideration for Pop~III stars, respectively. Cosmological simulation results (\citealt{lb2020}; LB20) are shown with the orange shaded region, with upper and lower limits denoted by filled triangles (for \textit{star forming} FOF haloes) and the dashed curve (for $\nu<4$ peaks), respectively.}
    \label{f2}
\end{figure}

The degeneracy of $\eta_{0}$ and $M_{\rm th}$ can be broken by further information on either one of them. For illustration, we consider the constraints on $M_{\rm th}$ from the cosmological simulations in \citet{Anna2018} with different levels of baryon-DM streaming motion (i.e. $v_{\rm stream}=0,\ 1,\ 2,\ 3\sigma$), and the upper limits on $\eta_{0}$ from abundance matching at $z\lesssim 2$ \citep{behroozi2019universemachine} and reionization (see the next subsection). If we fix $M_{\rm th}$ to the mass above which more than 50\% of haloes can form stars, using the results in \citet{Anna2018}, we can derive the redshift evolution of $\eta_{0}$ at different levels of streaming motion, as shown Fig.~\ref{f2}. We find rapid evolution of $\eta_{0}$ with redshift, which is insensitive to streaming motion (within a factor of 4). For instance, we have $\eta_{0}\simeq 0.012$, $6.2\times 10^{-3}$, $1.6\times 10^{-3}$ and $4\times 10^{-4}$ at $z_{1}\simeq 13.1$, $14.2$, $15.5$ and $17.3$, in the most representative case of $1\sigma$ streaming motion. These results can be fitted to an exponential form $\eta_{0}\simeq 2.7\times 10^{-3-0.37(z-15)}$. 

We then compare our results with the estimations of $\eta_{0}$ based on the EDGES 21-cm absorption signal \citep{nature21cm} in the semi-analytical models from \citealt{schauer2019constraining} (SA19) and \citealt{Mirocha2019} (MJ19), as well as the results from the cosmological simulation in \citealt{lb2020} (LB20; their fiducial run \texttt{FDbox\_Lseed}). 
For LB20, we consider two definitions of HSFE in simulations: (i) ratio of total increased stellar and halo masses for \textit{star forming} haloes, identified by the \textsc{rockstar}\footnote{\url{https://bitbucket.org/gfcstanford/rockstar/src/master/}} halo finder \citep{behroozi2012rockstar} based on the friends-of-friends (FOF) method, i.e. $\eta_{0}=\Delta \left(\sum_{j}M_{\star,j}\right)/\Delta\left(\sum_{j}M_{\mathrm{halo},j}\right),\ M_{\star,j}>0$; (ii) ratio of the increased \textit{simulated} stellar mass and increased mass in collapsed objects at $\nu<4$ peaks predicted by the EPS formalism\footnote{In detail, $\eta_{0}=\Delta M_{\star,\rm sim}/\Delta M_{\rm col}(\nu<4)$ with $M_{\rm col}(\nu<4)=V_{C}\int_{M_{\rm th}}^{M_{\rm crit}}Mn_{\rm h}(M)dM$ (as a function of redshift), where $V_{C}$ is the simulation volume, $n_{\rm h}(M)$ the EPS halo number density per unit mass, $M_{\rm crit}$ the critical mass for $\nu=4$ peaks ($\sigma(M_{\rm crit})=\delta_{\rm c}/4$), and $M_{\rm th}=1.63\times 10^{6}\ \mathrm{M_{\odot}}$ according to SA19. Note that the HSFE calculated in this way is lower than that defined for all simulated FOF haloes (i.e. $\eta_{0}=\Delta M_{\star,\rm sim}/\Delta\left(\sum_{j}M_{\mathrm{halo},j}\right),\ M_{\mathrm{halo},j}>M_{\rm th}$) by a factor of 2, reflecting the well-known discrepancies between the EPS halo mass functions and those measured in simulations.}. 

The extrapolated value of $\eta_{0}$ at $z\sim 20$ from our results agrees well with the estimations in SA19\footnote{\citet{Anna2018} also use the EPS formalism in their analysis, so that their definition of `halo' is consistent with the one adopted in this work.} and LB20 for $\nu<4$ peaks, showing that HSFE is typically low ($\sim$ a few $10^{-4}$) for Pop~III stars in minihaloes. 
This is required to not imprint the 21-cm absorption signal at a redshift ($z\gtrsim 20$) higher than observed. Note that similar values ($\sim 10^{-4}-10^{-3}$) are also found in the recent cosmological simulation from \citet{danielle2020}. 
However, in MJ19 and the \textit{star forming} FOF haloes of LB20, the HSFE is almost constant at a higher level ($\sim$ a few $10^{-3}$) at $z\sim 10-20$. The detailed analysis of semi-numerical simulations, taking into account the escape fraction of UV photons, also infers that $\eta_{0}\sim$ a few $10^{-3}$ from the observed 21-cm absorption signal \citep{fialkov2019,21cm2020}. In our case, $\eta_{0}\sim$ a few $10^{-3}$ at $z\sim 14-16$, but it rises to $\sim 10^{-2}$ at $z\sim 13$. The rapid evolution of $\eta_{0}$ may be caused by the fact that XLSSC is likely formed in an overdense region (i.e. at a $\nu\gtrsim 4$ peak). 

Note that for LB20, the HSFE based on \textit{star-forming} FOF haloes is higher by a factor of $\sim 3-9$ than that derived from the EPS formalism for $\nu<4$ peaks (see the orange shaded region in Fig.~\ref{f2}, representing the range in values for the two HSFE methods)\footnote{This implies that for low-mass haloes at high-$z$, where the delay time between halo and star formation is non-negligible, most haloes \textit{do not} host stars at a given snapshot. That is to say, the HSFE measured in \textit{star forming} haloes is not representative for the entire halo population, and must be diluted if the average should be taken over \textit{all} haloes.}. The discrepancies between SA19 and \citet{fialkov2019,21cm2020} may be caused by different estimations of host halo abundances and treatments for the escape fraction of Lyman-$\alpha$ photons. As significant uncertainties also exist in the input stellar age distribution of XLSSC~122 \citep{willis2020spectroscopic}, we do not expect the discrepancies found here to have statistical significance. Nevertheless, they indicate that caution is necessary when comparing the results from observations, semi-analytical models and simulations. 

\subsection{Upper limit from reionization}
\label{s3.2}
Another bound on the HSFE can be set by reionization, thus providing a consistency check for our analysis, and also constraining the underlying structure formation history (see the next subsection). We derive an upper limit on $\eta_{0}$ from reionization as follows. First, we define the completion of reionization as the moment when the number of ionizing photons per hydrogen atom reaches two \citep{so2014fully}, such that
\begin{align}
\frac{2X_{\uH}\bar{\rho}_{\rm b}}{m_{\uH}}&=\dot{N}_{\rm ion}t_{\star}\hat{f}_{\rm esc}\notag\\
&\times\int_{M_{\rm atm}}^{\infty}M\eta\left[n_{\rm h}(t_{\rm ion})-n_{\rm h}(t_{\rm ion}-t_{\star})\right]dM\ ,\label{e4}
\end{align}
where $X_{\uH}=0.76$, $\bar{\rho}_{\rm b}=3\Omega_{\rm b} H_{0}^{2}/(8\pi G)$ is the average baryon density, $\eta\equiv \eta(\eta_{0, z_{\rm ion}},M)$ the HSFE parameterized by $\eta_{0,z_{\rm ion}}$ at $z_{\rm ion}$, $M_{\rm atm}=2.5\times 10^{7}\ \mathrm{M_{\odot}} [(1+z_{\rm ion})/10]^{-3/2}$ the atomic cooling threshold, and $n_{\rm h}(t)\equiv n_{\rm h}(t,M)$ the halo mass function at time $t$. Further, $t_{\rm ion}$ is the age of the Universe at the end of reionization corresponding to $z_{\rm ion}$, $t_{\star}\sim 10$~Myr and $\dot{N}_{\rm ion}=10^{47}\ \mathrm{s^{-1}\ M_{\odot}^{-1}}$ are the lifetime of O/B stars and luminosity of ionizing photons per unit stellar mass for Population~II stars\footnote{Here we neglect the contribution of Pop~III stars to completing reionization at $z_{\rm ion}\sim 6$, as Pop~III star formation will be significantly suppressed by external Lyman-Werner and ionizing photons, as well as metal enrichment. Previous studies have found that the Pop~III contribution to the ionizing photon budget is $\lesssim 10\%$ (e.g. \citealt{greif2006two,wise2011birth,paardekooper2015first}). } (which dominate at $z\lesssim 18$, see e.g. \citealt{lb2020}), and $\hat{f}_{\rm esc}$ is the effective escape fraction. Then we assume that $\eta_{0,z_{\rm ion}}\ge \eta_{0}(z_{1})$ for $z_{1}\gtrsim 13$, such that the upper limit on $\eta_{0,z_{\rm ion}}$ to not complete reionization before $z_{\rm ion}$ is also the upper limit on $\eta_{0}(z_{1})$, which can be derived by equation~(\ref{e4}) given $\hat{f}_{\rm esc}$ and $z_{\rm ion}$. Here we adopt $z_{\rm ion}=6$, 
and consider the range $\hat{f}_{\rm esc}\sim 0.1-0.7$ with $\hat{f}_{\rm esc}=0.3$ the fiducial value, based on the simulations of \citealt{so2014fully} (see also \citealt{paardekooper2015first}).

As shown in Fig.~\ref{f1} and \ref{f2}, the inferred $\eta_{0}$ at $z_{1}\gtrsim 13$ is lower than the upper limits set by abundance matching and reionization for $v_{\rm stream}\le 2\sigma$, which accounts for $\approx 99\%$ of the cosmic volume, further demonstrating that our results are consistent with existing constraints. Overall, we confirm the emerging picture that star formation began early in cosmic history, but was initially quite inefficient, with a ramp up towards a late epoch of reionization.

\subsection{Fuzzy dark matter}
\label{s3.3}

Given $M_{\rm th}$ and the upper limit on $\eta_{0}$ either from abundance matching or reionization, our model can also place constraints on the underlying structure formation history (captured by $n_{\rm p}$ and $n_{\rm h}$), governed by DM physics. As an example, we consider the fuzzy dark matter (FDM) scenario, parameterized by the mass of ultra-light particles, $m_{\rm a}$, whose linear power spectrum is given by \citep{hu2000fuzzy}
\begin{align}
\begin{split}
P_{\rm FDM}(k)&=T_{\rm FDM}^{2}(k)P_{\rm CDM}(k)\ ,\\
T_{\rm FDM}(k)&=\cos [x_{\rm J}^{3}(k)]/[1+x_{\rm J}^{8}(k)]\ ,\\
x_{\rm J}(k)&=1.61(m_{\rm a}c^{2}/10^{-22}\ \mathrm{eV})^{1/18}(k/k_{\rm J,eq})\ ,\\
k_{\rm J,eq}&=9(m_{\rm a}c^{2}/10^{-22}\ \mathrm{eV})^{1/2}\ \mathrm{Mpc^{-1}}\ .
\end{split}\label{e5}
\end{align}
As shown in \citet{hirano2018first}, star formation can be significantly delayed to occur in more massive structures in FDM models, compared with standard $\Lambda$CDM. Therefore, the mass of old stars in XLSSC~122 can constrain the parameter $m_{\rm a}$, once $\eta_{0}$ and $M_{\rm th}$ are known. For simplicity, we now consider all the redshift bins together\footnote{Actually $\eta_{0}$ here stands for the \textit{cumulative} HSFE for $z\gtrsim z_{1}\simeq 12.7$. We assume that the HSFE generally increases with decreasing redshift before reionization, so that the cumulative HSFE should be smaller than the instantaneous one at a given redshift. Therefore, our constraints on $m_{\rm a}$ should be regarded conservative when any upper limit on the \textit{instantaneous} HSFE is adopted.} for stars/galaxies older than 2.98~Gyr, which accounts for $f_{\rm old}\simeq 0.43$ of the total stellar mass, and rewrite equation~(\ref{e3}) at $z_{1}\simeq 12.7$ as
\begin{align}
M_{\star,\rm old}=\int_{M_{\mathrm{th}}}^{M_{2}}n_{\rm p}(M_{1})M_{1}\eta(\eta_{0},M_{1})dM_{1}\ ,\ \label{e6}
\end{align}
where $n_{\rm p}(M_{1})\equiv n_{\rm p}(z_{1},M_{1}|z_{2},M_{2})$ (see Fig.~\ref{f4}), and $M_{\star,\rm old}=f_{\rm old}M_{\star,\mathrm{tot}}$. 
There exists a lower limit $m_{\rm a,\min}$ below which the above equation~(\ref{e6}) cannot be satisfied with reasonable $M_{\rm th}$ and $\eta_{0}$, when structure formation is delayed to lower redshifts $z\lesssim z_{1}\simeq 12.7$. 

\begin{figure}
    \centering
    \includegraphics[width=\columnwidth]{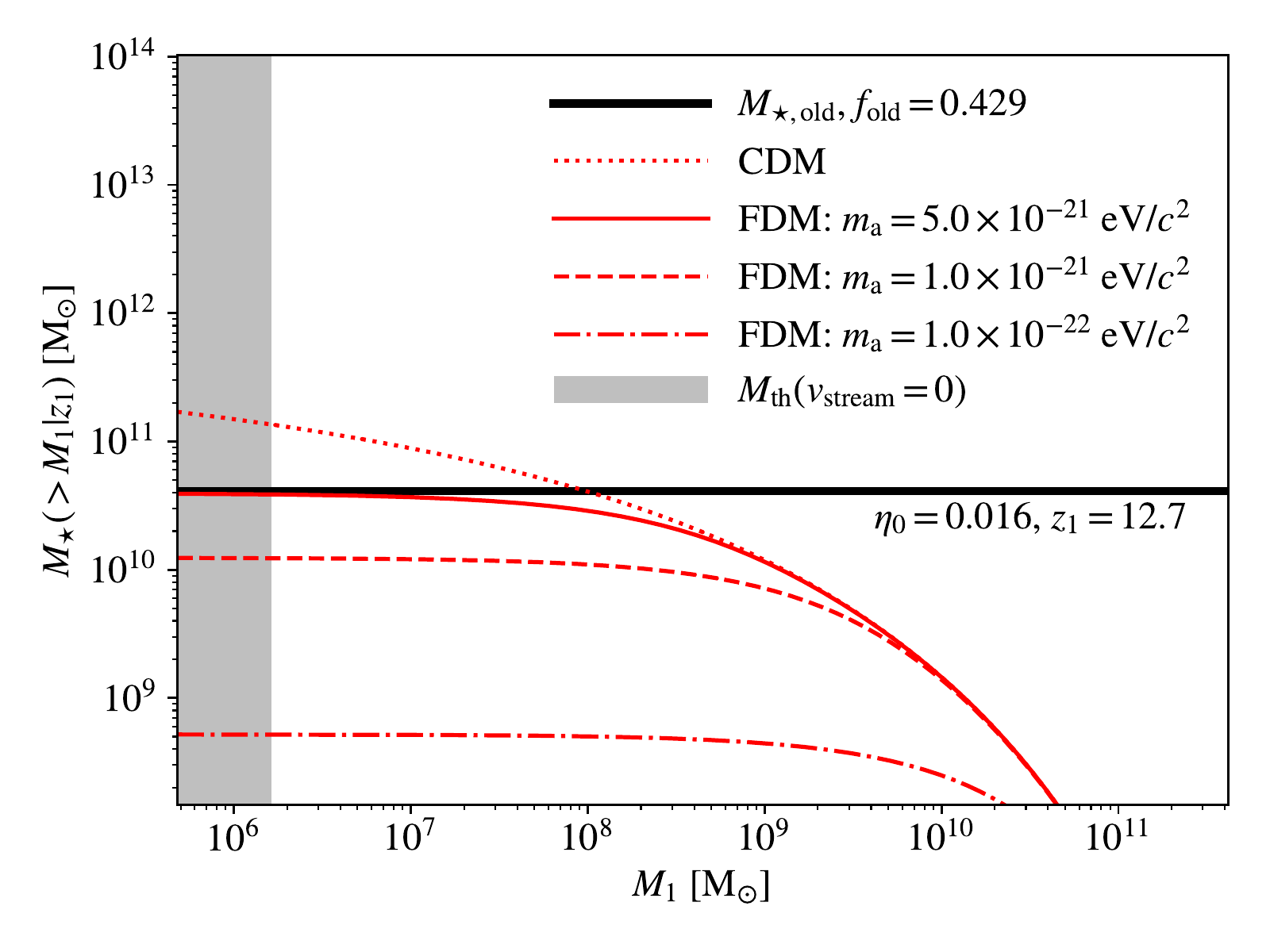}
    \caption{Cumulative stellar mass from the progenitors of XLSSC~122 at $z_{1}= 12.7$, for the standard $\Lambda$CDM (dotted), and FDM models with $m_{\mathrm{a}}c^{2}=5\times 10^{-21}\ \mathrm{eV}$ (solid), $10^{-21}\ \mathrm{eV}$ (dashed) and $10^{-22}\ \mathrm{eV}$ (dash-dotted). Here $\eta_{0}=0.016$ is set to the fiducial upper limit (with $\hat{f}_{\mathrm{esc}}=0.3$) placed by reionization at $z_{\rm ion}= 6$ \citep{so2014fully} for the critical FDM model with $m_{\mathrm{a}}c^{2}=5\times 10^{-20}\ \mathrm{eV}$. The total mass of old galaxies (older than 2.98~Gyr, i.e. formed at $z>z_{1}$) in XLSSC~122 is shown with the thick horizontal line, which makes up $f_{\mathrm{old}}\simeq 0.43$ of the total stellar mass. The shaded region shows the range of halo mass in which up to 50\% of haloes have star formation from \citet{Anna2018} without streaming motion. The FDM models that never cross the $M_{\star,\rm old}$ line are ruled out.}
    \label{f3}
\end{figure}

To derive a conservative estimate for this lower limit, we set $M_{\rm th}$ to the lowest threshold value of $\simeq 5\times 10^{5}\ \mathrm{M_{\odot}}$, which is the minimum mass of star formation in \citet{Anna2018} with no streaming motion\footnote{It turns out that $m_{\rm a,\min}$ is not sensitive to $M_{\rm th}$ for $M_{\rm th}\lesssim 10^{7}\ \rm M_{\odot}$ (i.e. $v_{\rm stream}\lesssim 2\sigma$), since for typical FDM models with $m_{\rm a}\lesssim 10^{-20}\ \mathrm{eV}/c^{2}$, the contribution from minihaloes ($M_{\rm halo}\lesssim 10^{7}\ \rm M_{\odot}$) to the total stellar mass formed at $z\gtrsim z_{1}\simeq 12.7$ is negligible (see Fig.~\ref{f3}).}, and $\eta_{0}$ to the (fiducial) upper limit from reionization $\eta_{0,\max}$ (with $\hat{f}_{\rm esc}=0.3$), given by equation~(\ref{e4}). Note that here both the halo mass function $n_{\rm h}$ and the progenitor mass function $n_{\rm p}$ depend on $m_{\rm a}$, so that equations~(\ref{e3}) and (\ref{e5}) must be solved together for $\eta_{0,\max}$ and $m_{\rm a,\min}$. Carrying out these steps, we find $\eta_{0,\max}\simeq 0.016$ and $m_{\rm a,\min}\simeq 5\times 10^{-21}\ \mathrm{eV}/c^{2}$. 

This approach is further illustrated in Fig.~\ref{f3} in terms of the cumulative stellar mass $M_{\star}(>M_{1})$, where FDM models with $M_{\star}(>M_{\rm th})<M_{\star,\rm old}$ are ruled out. 
If we adopt $\eta_{0}=0.025$, which is the local ($z\lesssim 2$) upper limit from abundance matching \citep{behroozi2019universemachine}, equation~(\ref{e6}) can be solved independently to give a weaker constraint on FDM of $m_{\rm a,\min}\simeq 2.5\times 10^{-21}\ \mathrm{eV}/c^{2}$. Using the reionization upper limit with $\hat{f}_{\rm esc}=0.1$ leads to an even weaker constraint of $m_{\rm a,\min}\simeq 6\times 10^{-22}\ \mathrm{eV}/c^{2}$ for $\eta_{0,\max}\simeq 0.091$, which is unphysical, as it would imply that $\sim$60\% of baryons end up in stars. Interestingly, our constraints on FDM inferred from the cluster XLSSC~122 are consistent with those based on the EDGES 21-cm absorption signal \citep{nature21cm}, e.g. $m_{\rm a,\min}\simeq 5\times 10^{-21}\ \mathrm{eV}/c^{2}$ in \citet{lidz2018} and $m_{\rm a,\min}\simeq 8\times 10^{-21}\ \mathrm{eV}/c^{2}$ in \citet{schneider2018}.

\section{Summary and Conclusions}
\label{s4}
We demonstrate a new approach of indirectly constraining early star and structure formation via mature galaxy clusters at cosmic noon ($z\sim2$), using the cluster XLSSC~122 as an example ($z_{\rm obs}=1.98$). Based on the age distribution of galaxies/stars in XLSSC~122 (neglecting dust extinction) measured by HST photometry \citep{willis2020spectroscopic}, and its halo properties from X-ray observations \citep{mantz2018xxl}, we infer a rapid evolution of the halo star formation efficiency (HSFE, $\eta\equiv\langle\Delta M_{\star}\rangle/\Delta M_{\rm halo}$) at $z\sim 13-18$. Specifically, we derive a fit $\eta\simeq 2.7\times 10^{-3-0.37(z-15)}$ for low-mass haloes ($M_{\rm halo}\lesssim 10^{10}\ \rm M_{\odot}$) that host the first stars and galaxies. Our results generally agree with semi-analytical models based on 21-cm absorption and cosmological simulations, giving $\eta_{0}\sim 10^{-4}$ to a few $10^{-3}$ at $z\sim 13-20$ \citep{Mirocha2019,schauer2019constraining,fialkov2019,21cm2020,danielle2020,lb2020}. However, such rapid evolution is unique to our model, likely caused by the fact that XLSSC is formed in an overdense region, corresponding to a $\nu\gtrsim 4$ peak. 

We also place new constraints on the mass of ultra-light bosons in fuzzy dark matter models of $m_{\rm a}\lesssim 5\times 10^{-21}\ \mathrm{eV}/c^{2}$, from the abundance of star forming galaxies at $z\gtrsim 13$ in the merger tree of XLSSC~122. 
This is comparable to existing constraints $m_{\rm a}\lesssim 5-8\times 10^{-21}\ \mathrm{eV}/c^{2}$ \citep{lidz2018,schneider2018}. 

However, significant uncertainties exist in the inferred stellar age distribution of XLSSC~122, as the posterior age distributions of individual galaxies are broad ($\sim \rm Gyr$). This is typical for photometry-inferred ages (see fig.~7-8 in \citealt{andreon2014jkcs} and extended data fig.~2-3 of \citealt{willis2020spectroscopic}). The age distribution is also sensitive to the underlying stellar population parameters assumed for SED fitting (e.g. IMF, metallicity and star formation history), especially the assumption on dust absorption \citep{willis2020spectroscopic}. We here focus on the zero-dust absorption ($A_{V}=0$) model that predicts major star formation to occur around $z\sim 12$ and extend to $z\sim 18$, while in models with dust absorption (e.g. $A_{V}=0.3$ and $0.5$) the stellar population is shifted to lower redshifts ($z\sim 6-13$), becoming less relevant to the first stars and galaxies. 

Besides, it remains unknown whether XLSSC~122 is an extreme case or a typical galaxy cluster at $z\sim 2$. Therefore, our results should be regarded as tentative and for illustration purpose. Nevertheless, more comprehensive results will be obtained if our approach is extended to a large sample of clusters or field post-starburst galaxies at cosmic noon, with a full statistical framework in which observational uncertainties are properly propagated to the inferred star/structure formation parameters. There is thus a strong case for systematic observational campaigns to identify galaxy clusters at cosmic noon and to characterize their member galaxies. 

On the theoretical side, we use the standard EPS formalism for simplicity and flexibility in the current work, which defines haloes differently from cosmological simulations. This introduces an additional layer of complexity for bridging theory and observation, given that simulations are needed to implement the detailed physics of star and galaxy formation, such as primordial chemistry, cooling and stellar feedback. In future work, merger trees constructed from simulations should be used to calculate the progenitor mass functions of clusters. That approach does not only remove the ambiguity in halo definition but also enables one to account for cosmic variance. We may also use more physically motivated models of the HSFE that allow variation with halo mass at $M_{\rm halo}\lesssim 10^{10}\ \rm M_{\odot}$, reflecting the different modes of early star formation in atomic cooling haloes ($M_{\rm halo}\gtrsim 10^{7}\ \rm M_{\odot}$) and molecular cooling minihaloes ($M_{\rm halo}\sim 10^{6}\ \rm M_{\odot}$).

Overall, we begin to probe the earliest epoch of star and galaxy formation with the tantalizing hints provided by pioneering observations, such as the ones discussed here. Soon, we will be able to complement this with direct observations of active star formation at the highest redshifts, together contributing to the emerging model of the first stars and galaxies. 

\section*{Acknowledgments}
Support for this work was provided by NASA through the NASA Hubble Fellowship grant HST-HF2-51418.001-A awarded  by  the  Space  Telescope  Science  Institute,  which  is  operated  by  the Association  of  Universities for  Research  in  Astronomy,  Inc.,  for  NASA,  under  contract NAS5-26555.

\bibliographystyle{mnras}
\bibliography{ref} 


\bsp	
\label{lastpage}

\end{document}